\documentclass{osa-article}

\journal{oe}

\usepackage[utf8]{inputenc}
\usepackage[T1]{fontenc}
\usepackage{xcolor,soul}
\usepackage{color, colortbl}
\usepackage{xfrac}

\articletype{Research Article}

\begin{document}

\title{Realization and simulation of high power holmium doped fiber laser for long-range transmission}

\author{Julien Le Gouët,\authormark{1,*} François Gustave,\authormark{1} Pierre Bourdon,\authormark{1} Thierry Robin,\authormark{2} Arnaud Laurent,\authormark{2} and Benoit Cadier \authormark{2}}

\address{\authormark{1}Office National d'Etudes et de Recherches Aérospatiales (ONERA), Palaiseau, France\\
\authormark{2}iXblue Photonics, Lannion, France\\}

\email{\authormark{*}julien.le\_gouet@onera.fr } 

\begin{abstract}
We report on our realization of a high power holmium doped fiber laser, together with the validation of our numerical simulation of the laser. We first present the rare absolute measurements of the physical parameters that are mandatory to model accurately the laser-holmium interactions in our silica fiber. We then describe the realization of the clad-pumped laser, based on a triple-clad large mode area holmium (Ho) doped silica fiber. The output signal power is 90~W at 2120~nm, with an efficiency of about 50\% with respect to the coupled pump power. This efficiency corresponds to the state of the art for clad-pumped Ho-doped fiber lasers in the 100~W power class. By comparing the experimental results to our simulation, we demonstrate its validity, and use it to show that the efficiency is limited, for our fiber, by the non-saturable absorption caused by pair induced quenching between adjacent holmium ions.
\end{abstract}

\section{Introduction}
In various applications like long-range free-space optical telecommunications, wind lidar, or defense systems, the requirements of the laser source are the following: robustness to harsh environments, high average power (>100W), low thermal load, single spatial mode (M$^2\simeq 1$), high transmission through the atmosphere, eye-safe radiation, and high electro-optic efficiency. 
When it comes to conciliate all these requirements, all-fiber laser sources appear as the best compromise.

Several solutions can then be considered. For example, the fiber telecommunication band offers some atmospheric transmission windows which can be addressed with an erbium-doped fiber laser, emitting between 1560~nm and 1580~nm and pumped in-band by laser diodes at 1532~nm to avoid large quantum defect \cite{Zhang_2011}. However, infrared laser diodes suffer an efficiency drop at high power, due to various thermally activated loss mechanisms \cite{Liang_2011}. A thulium (Tm) fiber laser at about 2040~nm should also be able to meet the requirements above, by using laser diode pumping at 793~nm \cite{Goodno_2009, Ehrenreich_2010}, but thermal load is expected to limit the power scaling. In-band pumping of a Tm fiber by another Tm fiber laser is an alternative, at the expense of the efficiency for laser wavelengths longer than 2050~nm \cite{Sincore_2018}. In the long term perspective of scaling the power up to several kW, another promising option consists in pumping holmium (Ho) doped fibers in the $^5I_8-{^5I_7}$ transition (close to 1950~nm) with a Tm fiber laser. 


Since the record power of 400~W and 50\% efficiency established with an all-fiber clad pumped configuration \cite{Hemming_2013}, this type of laser source attracted interest from a number of other laboratories \cite{Friebele_2014,Aubrecht_2016,Sincore_2016,Shardlow_2016,Holmen_2018}. It should be noted that other configurations can allow higher efficiency in the class of 100~W laser power (e.g. core-pumped \cite{Baker_2018} or double-end pumped \cite{Hemming_double_end_2013}, but these configurations are not compatible with the requirements of low thermal load or robustness.

En route to a high power laser source based on in-band pumped Ho fibers, one of the limitations to the efficiency lies in the silica purity, for clad and core, in terms of hydroxyl OH concentration. Indeed these impurities, whom concentration can vary with the fabrication method, present absorption bands in the spectral range around 2~\textmu m which includes pump and signal wavelengths \cite{Humbach_1996}. Another limitation to the laser efficiency in highly doped fibers stems from the pair-induced quenching, which translates at high power into a non-saturable absorption. Both contributions to the pump and signal attenuations must be precisely measured and taken into account.

In this article we resume our work, from the spectroscopy of the holmium ions in single-mode fiber (SMF) samples to the realization of a high efficiency high power laser based on a triple-clad large mode area (LMA) fiber. We compare these results to our numerical simulation of the fiber laser, based on all the physical parameters measured on the SMF samples and on the LMA fiber. We finally discuss the main limitations of efficiency in the case of our fiber laser.

\section{Spectroscopy of Ho in a single-mode sample}
\label{Sec_spectro}

The generation of a laser signal in a rare-earth doped fiber cavity is ruled by a set of very well-known equations for population rate and propagation \cite{Giles_1991}. To solve these equations, the following parameters must be determined: linear absorption by the active ion, cross-section spectra for absorption and emission, excited state lifetime, and propagation passive loss. In addition, high holmium concentrations allow the formation of ion pairs which can reduce the laser efficiency through non-radiative energy transfers \cite{Wagener_1993,Kurkov_2010}, so the fraction of ion pairs must be measured. The values of all these physical parameters can vary significantly with the composition of the fiber, so they must be measured in detail to permit an accurate simulation of the fiber laser.

For rare-earth ions in silica fibers, the principle of measurement of each of these parameters is quite universal and already well documented in the literature. Therefore the principle and setup are only briefly exposed. Our results are commented and compared to previous measurements, when possible.

\subsection{Absorption cross-section spectra}
Some values of the absorption cross-section $\sigma_\text{abs}$ have been already published for the $^5I_8-{^5I_7}$ transition of holmium in silica fibers. However this value can vary significantly with the co-dopant ratios in the core, different from one fabrication process to another. Subsequently, there is a strong dispersion of the peak value of $\sigma_\text{abs}$ in the literature for Ho doped silica fibers. Here are some values of $\sigma_\text{abs}$ at 1950~nm that are given or can be infered: $2.9 \ 10^{-25}\text{m}^2$ \cite{Simakov_2013}, $4.65 \ 10^{-25}\text{m}^2$ \cite{Wang_2015}, $5.8 \ 10^{-25}\text{m}^2$ \cite{Simakov_2016}, and up to $8.5 \ 10^{-25}\text{m}^2$ \cite{Ryabochkina_2015}. A specific measurement on our own samples seems thus necessary.

The absorption cross-section at a given wavelength $\sigma_\text{abs}(\lambda)$ is obtained from the holmium concentration in the fiber core and from the linear absorption at $\lambda$, following the expression 
\begin{equation}\label{eq_absorption}
\alpha_\text{Ho}(\lambda)=\sigma_\text{abs}(\lambda).[\text{Ho}].\Gamma(\lambda)
\end{equation}
where $\Gamma(\lambda) = 1-\text{exp}\left[-2.\sfrac{w_\text{core}^2(\lambda)}{w_\text{mode}^2(\lambda)}\right]$ corresponds to the overlap integral between the fundamental mode (beam radius $w_\text{mode}$) and the active core (radius $w_\text{core}$). 

The linear absorption spectrum is obtained by the usual cut-back method \cite{Desurvire_EDFA} from a SMF single-clad sample (iXblue HDF-PM-8-130). At the absorption peak wavelength $\lambda_0=1946$~nm, this moderately doped fiber (defined here as fiber \#1) presents an absorption of $57\pm 1$~dB/m. The core composition of this SMF sample was measured with a scanning electron microscope equipped for wavelength dispersive spectroscopy (WDS), and revealed a Ho$^{3+}$ concentration $[\text{Ho}]=(3.7 \pm 0.6)10^{25}\ \text{m}^{-3}$. The fiber has a 8~\textmu m core diameter and a numerical aperture NA=0.17, so the mode diameter at $\lambda_0$ is $w_\text{mode}=9.4$~\textmu m, and the overlap parameter is $\Gamma(\lambda_0)=0.76$. We could thus infer an absorption cross-section $\sigma_\text{abs}(\lambda_0)=(4.8 \pm 0.8)10^{-25}\ \text{m}^2$.

\begin{figure}[h!]
\centering\includegraphics[width=12cm, trim={.0cm 0cm 0 0}, clip]{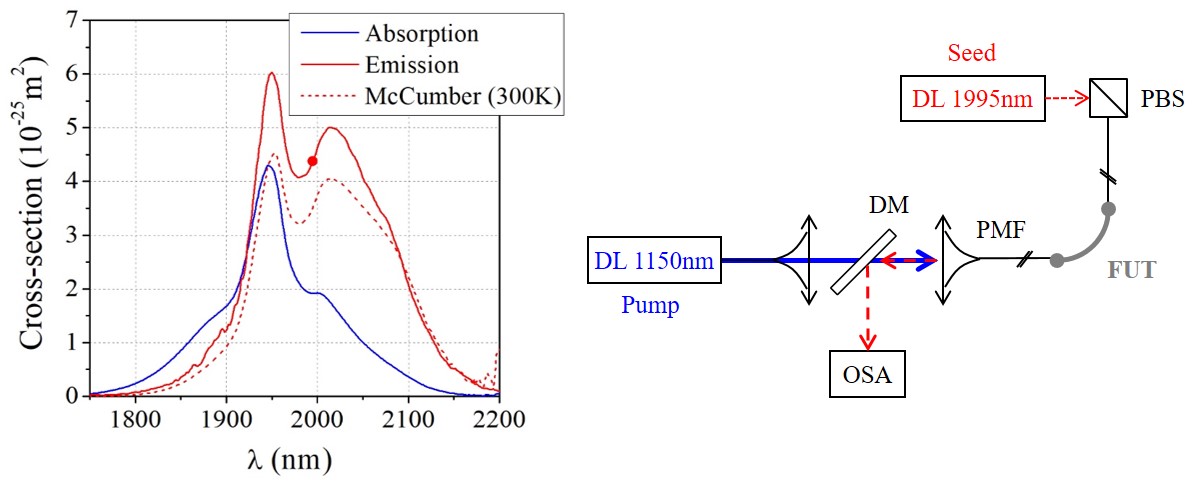} 
\caption{Left: cross-section spectra for absorption (blue line) and emission (red line) of the Ho doped single mode fiber. The emission spectrum calculated from McCumber equation is displayed in dotted line. The red dot on the emission spectrum indicates the position $\lambda_\text{test}=1995$~nm where the small-signal gain was measured. Right: setup for the measurement of the small-signal gain (seed on) or the fluorescence spectrum (seed off).}
\label{Fig_Ho_spectra_setup}
\end{figure}

Another HDF-PM-8-130 sample (fiber \#2) was prepared with a higher Ho$^{3+}$ concentration. The relative shape of the absorption spectrum is identical to that of the previous sample. The linear absorption of this fiber is $425\pm5$~dB/m at 1950~nm, with the same 8~\textmu m core diameter and NA=0.21 ($\Gamma(\lambda_0)=0.85$). The WDS analysis yields a Ho$^{3+}$ concentration $[\text{Ho}]=(27 \pm 1) 10^{25} \ \text{m}^{-3}$. Thanks to the better sensitivity on concentration measurement, the estimation of the absorption cross-section is narrowed down to $\sigma_\text{abs}(\lambda_0)=(4.3 \pm 0.2) 10^{-25}\ \text{m}^2$ (see Fig.~\ref{Fig_Ho_spectra_setup}, left). 

\subsection{Emission cross-section spectrum}

The absolute measurement of the emission cross-section spectrum $\sigma_\text{em}(\lambda)$ is not straightforward in general and we present here, to the best of our knowledge, its first realization on a holmium doped fiber. The most accurate method to obtain $\sigma_\text{em}(\lambda)$ consists in measuring the relative shape $S_\text{fluo}(\lambda)$ of the fluorescence spectrum, and determining the absolute emission cross-section at a particular wavelength to scale the whole spectrum.

In homogeneously broadened transitions such as those involved in the rare-earth doped fiber lasers at room temperature, all the sublevels of all the ions in the excited state contribute to the emission. To make sure that all the possible contributions are taken into account in the fluorescence and gain measurements, the ions must be pumped to an energy level higher than the excited laser state \cite{Desurvire_EDFA}. For the emission cross-section from the $^5I_7$ level, it is thus necessary to pump the ions into the $^5I_6$ level or higher. In a silica fiber, the $^5I_8-{^5I_6}$ transition corresponds to a wavelength of about 1150~nm \cite{Jackson_2007}.

The setup used to measure the small-signal gain and the fluorescence spectrum is illustrated on Fig.~\ref{Fig_Ho_spectra_setup} (right). The population inversion is provided by a 1150~nm laser diode that delivers 400~mW on a linear polarization. This pump beam is collimated, transmitted through a dichroic mirror (DM) and coupled into the core of the fiber under test (FUT=fiber \#1), which is polarization maintaining (PM). At the FUT output, a fiber coupled polarizing beam splitter (PBS) allows the injection of the small signal, propagating in the opposite direction of the pump, with an orthogonal polarization. The small signal itself consists of about $1~\mu$W at $\lambda_\text{test}=$1995~nm, delivered by a PM fiber coupled laser diode. The fluorescence or amplified signal is redirected by the DM towards an optical spectrum analyzer (OSA).

The length of the FUT must be chosen such that the population is completely inverted along the sample. A rule of thumb suggests that the sample should be much shorter than the inverse of the holmium absorption $\alpha_\text{Ho}(1150\ \text{nm})$. From \cite{Kurkov_2010} we find $\alpha_\text{Ho}(1950\ \text{nm})/\alpha_\text{Ho}(1150\ \text{nm})\simeq 2$, so for the moderately doped sample (fiber \#1) we estimate $\alpha_\text{Ho}(1150\ \text{nm})\simeq 29~\text{dB/m}\simeq 7~\text{m}^{-1}$. Thus the FUT should be much shorter than 14~cm to guarantee a complete population inversion along the sample. By shortening the fiber sample from 10~cm to 5~mm, and acquiring the corresponding fluorescence spectra, we find that they become identical for $L_\text{FUT}\lesssim2$~cm. After correction from the transmission spectrum between the FUT and the OSA (lenses, DM, fiber coupling...), we obtain a relative emission cross-section spectrum.

In order to determine the absolute emission cross-section, we measure the small-signal gain at one particular wavelength $\lambda_\text{test}$. Indeed, the ratio between absorption and emission cross-sections $\sigma_\text{em}/\sigma_\text{abs}$ at a given wavelength $\lambda$ is equal to the ratio $g^*/\alpha_\text{Ho}$ between the corresponding small-signal absorption and gain at $\lambda$ \cite{Desurvire_EDFA}. Then the peak emission cross-section of the $^5I_7-{^5I_8}$ transition is obtained by the following identity:
\begin{eqnarray}
\label{eq_CS_em_abs}
\dfrac{\sigma_\text{em}^\text{peak}}{\sigma_\text{abs}^\text{peak}}&=&  \dfrac{\sigma_\text{em}^\text{peak}}{\sigma_\text{em}(\lambda_\text{test})}.\dfrac{\sigma_\text{em}(\lambda_\text{test})}{\sigma_\text{abs}(\lambda_\text{test})}.\dfrac{\sigma_\text{abs}(\lambda_\text{test})}{\sigma_\text{abs}^\text{peak}}
=\dfrac{S_\text{fluo}^\text{peak}}{S_\text{fluo}(\lambda_\text{test})}.\dfrac{g^*(\lambda_\text{test})}{\alpha_\text{Ho}(\lambda_\text{test})}.\dfrac{\alpha_\text{Ho}(\lambda_\text{test})}{\alpha_\text{Ho}^\text{peak}} \\ 
&=&\dfrac{S_\text{fluo}^\text{peak}}{S_\text{fluo}(\lambda_\text{test})}.\dfrac{g^*(\lambda_\text{test})}{\alpha_\text{Ho}^\text{peak}}
\end{eqnarray}
where the first ratio is obtained from the measurement of a fluorescence spectrum and yields $S_\text{fluo}^\text{peak}/S_\text{fluo}(\lambda_\text{test})= 1.37$.

The small-signal gain is obtained by comparing the powers at the input and output of the FUT, and taking into account the transmission of all the components or splices that are not common to the two measurements. On a fully inversed length of $20$~mm, we measure a gain of $1.16\pm 0.05$~dB, which corresponds to a linear small-signal gain $g^*(\lambda_\text{test}=1995\text{nm})= 58\pm3$~dB/m. By coincidence this value is very similar to the absorption at 1946~nm ($\alpha_\text{Ho}^\text{peak}=57\pm 1$~dB/m), and we find the scaling ratio $\sigma_\text{em}^\text{peak}/\sigma_\text{abs}^\text{peak}=1.40\pm 0.06$. Given the value of the peak absorption cross-section, we finally estimate the peak emission cross-section to $\sigma_\text{em}^\text{peak} = 6.0\pm 0.4  \ 10^{-25} \text{m}^2$. 

The difference between the peak emission and absorption may seem important, since most of the papers on Ho$^{3+}$-doped fibers report equal amplitudes at the peak \cite{Bowman_2013,Simakov_2013,Friebele_2014,Fan_2015}. However these authors calculate the emission cross-section using the well-known McCumber relation, and assuming that the absorption and emission spectra are equal at the absorption peak. Two reservations can be expressed against this method, justifying our effort for an absolute measurement.

First the use of the McCumber relation requires an accurate knowledge of the partition function in the ground and excited states of the laser transition \cite{Digonnet_2002}. A common method consists in assuming an even splitting between the energy sublevels in these states, and the energy splitting is obtained by dividing the spectral width of the homogeneously broadened transition by the number of sublevels \cite{Miniscalco_1991}. However, unlike in Kramers ions (e.g. Er$^{3+}$ or Yb$^{3+}$), the degeneracy of the sublevels of the non-Kramers ions (e.g. Ho$^{3+}$ and Tm$^{3+}$) can stay high in a host exhibiting a high degree of symmetry (like amorphous silica). In that case the numbers of sublevels to consider depends on the host, so their determination seems hazardous.

Second, the application of the McCumber relation requires that the linewidths $\Delta E_\text{ij}$ of each transition between a Stark sublevel \textit{i} of the lower manifold and a sublevel \textit{j} in the upper manifold satisfies the following condition \cite{McCumber_1964,Florea_1999}:
\begin{equation}
\label{eq_McCumber_cond}
\Delta E_\text{ij} \ll k_\text{B}T
\end{equation}
where $k_\text{B}$ is the Boltzmann constant and $T$ is the temperature. At room temperature and a central wavelength of 1950~nm, this corresponds to the condition $\Delta \lambda_\text{ij} \ll 80$~nm. Considering for example the zero-phonon transition of the absorption spectrum (see Fig.~\ref{Fig_Ho_spectra_setup}, left), we find a linewidth of about 60~nm. Therefore the condition above is not strictly respected, and the accuracy of the resulting emission cross-section spectra is questionable.

As an illustration, we plot on Fig.~\ref{Fig_Ho_spectra_setup} (left, dotted line) the emission cross-section $\sigma_\text{em}(\lambda)$ calculated with the McCumber relation from $\sigma_\text{abs}(\lambda)$, and assuming that the two spectra intersect at the absorption peak wavelength 1946~nm. The figure does illustrate a noticeable difference close to the peak, but also that the measured and calculated spectra almost overlap beyond 2100~nm. This probably explains why the McCumber calculation of the emission cross-section spectra do not totally alter the accuracy of the laser simulations, at least for $\lambda>2100$~nm. Eventually, let us remind that our value of emission cross-section seems to be the first report from a direct measurement, instead of indirect calculations.


\subsection{Laser excited state lifetime}
In principle, the setup for the measurement of the $^5I_7$ level lifetime is very similar to that of the emission cross-section \cite{Desurvire_EDFA}. However, the timescale of the fluorescence decay (on the order of 1~ms) requires a detection system with a faster rise time than the OSA, hence a smaller detection area, and a lower signal intensity. The power delivered by the 1150~nm laser diode appeared too low compared to the detection noise of our photodiode, so we realized a Yb-doped fiber laser, based on the design described in \cite{Miao_2014}. We then managed to pump a 10~cm sample of fiber \#1 with a power up to 1~W.

\begin{figure}[h!]
\centering\includegraphics[width=10cm, trim={.5cm 0cm 0 0}, clip]{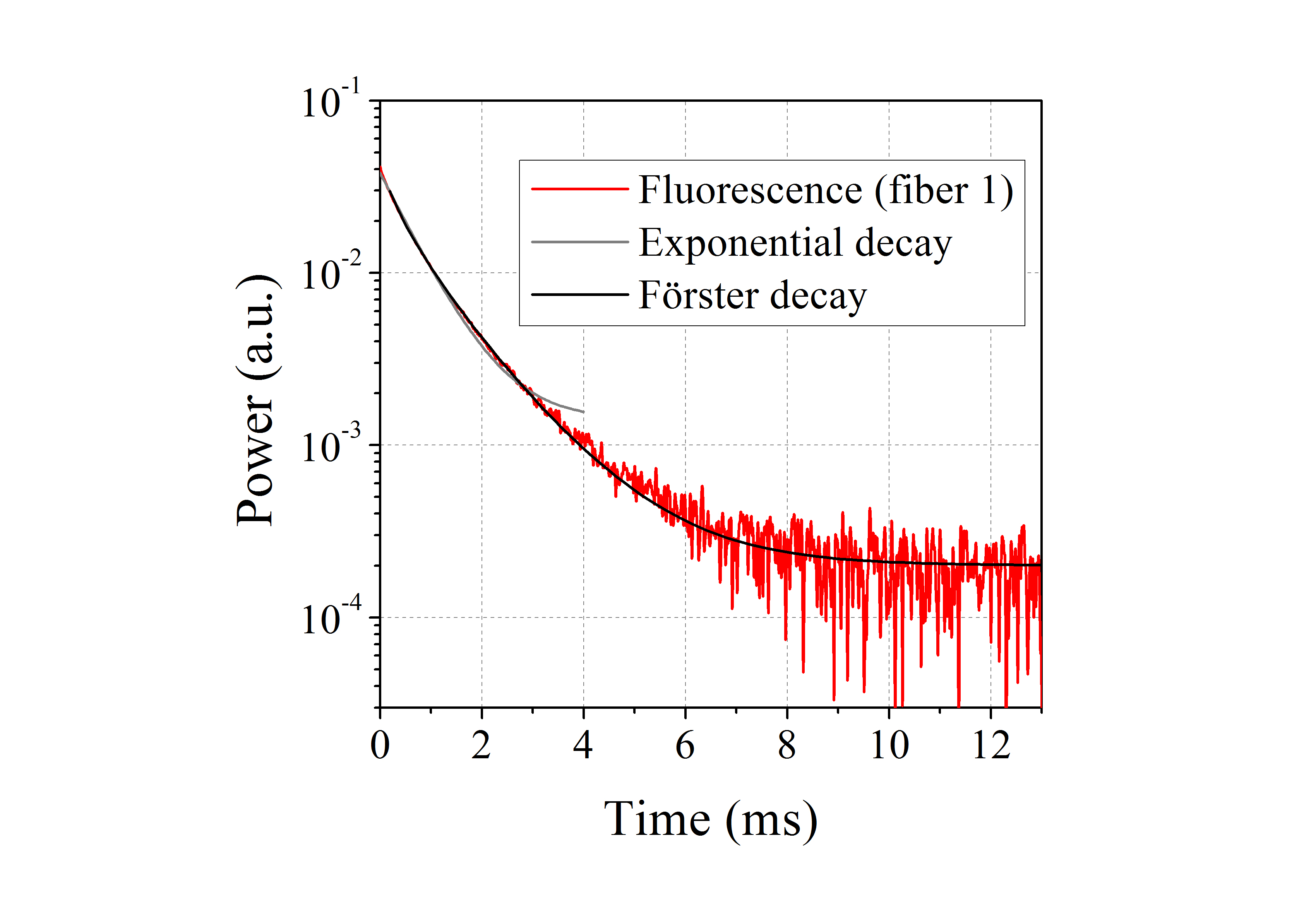} 
\caption{Temporal decay of the fluorescence from the $^5I_7$ excited level after optical pumping in the $^5I_6$ level. The experimental curve is fitted with two functions: the common single exponential decay (gray), as well as the non-exponential Förster decay function (black). The better fit of that last function illustrates the onset of energy transfers between adjacent excited holmium ions.}
\label{Fig_Ho_lifetime}
\end{figure}

In order to alternate population inversion and observation of fluorescence decay, we apply a 100~Hz square modulation on the pump power by adding a free-space acousto-optical modulator (AOM) on the path of the 1150~nm beam. In absence of Ho doped fiber, we find that the detection response time is about 10~\textmu s. The result of an averaging with the FUT is illustrated on Figure~\ref{Fig_Ho_lifetime}. The non-exponential decay can be a sign of energy transfers between adjacent Ho$^{3+}$ ions, typically through dipolar coupling \cite{Gomez_2016}. In that case the experimental curve can be well described by a Förster decay function \cite{Svelto_2010} : 
\begin{equation}
\label{eq_Forster_decay}
P(t)=P_0 + A. \text{exp}\left( -\gamma \sqrt{t} - t/\tau_0 \right) 
\end{equation}
and the fit yields the following parameters: $\tau_0=1.9$~ms and $\gamma=34\ \text{s}^{-1/2}$. A detailed modeling of the influence of the energy transfers between adjacent ions is out of the scope of this study. We thus let aside the modification of the rate equations, and consider that the decay of the excited state is described by an effective lifetime.

Fitting the first part of the fluorescence signal (until about 4~\textmu s) with a single exponential decay yields a value of $0.7\pm 0.1$~ms, which agrees with the value generally reported \cite{Hanna_1989,Bowman_2013}. The agreement between measurements and simulation, illustrated in the next sections, tends to suggest that this approximation does not alter notably the accuracy of the model.

\subsection{Background loss and hydroxyl concentration}
\label{Sec_mesureOH}

The attenuation at pump and signal wavelengths is a critical parameter for fiber sources, especially in a clad pumping configuration where the fiber can extend to more than 10~m to absorb the pump. In the spectral domain close to 2~\textmu m, where our pump and signal beams lie, the hydroxyl groups OH bound to the silica network present absorption peaks due to rovibrational transitions. A crucial concern in the realization of holmium doped optical fibers must be the reduction of the OH concentration, to optimize the laser efficiency.

The Holmium doped core material was manufactured using the solution doping technology. The process consists in soaking a porous silica layer with a water based solution containing dissolved Holmium and aluminum ions. However it is possible to obtain a low OH content by taking appropriate precautions. The porous silica layer must be deposited using a well maintained MCVD equipment fitted with high purity precursors, and with specific process well known from the fiber manufacturers. Low OH content silica, in such conditions, can be obtained routinely below 50 ppb. Once the porous layer has been soaked and the solution drained away, the subsequent drying procedure must include a chlorination step at high temperature  to remove any moisture trace, prior to vitrification and collapse. Provided this drying step is well designed, no additional OH content is observed.

\begin{figure}[h!]
\centering\includegraphics[width=10cm, trim={0cm 1.5cm 0 0}, clip]{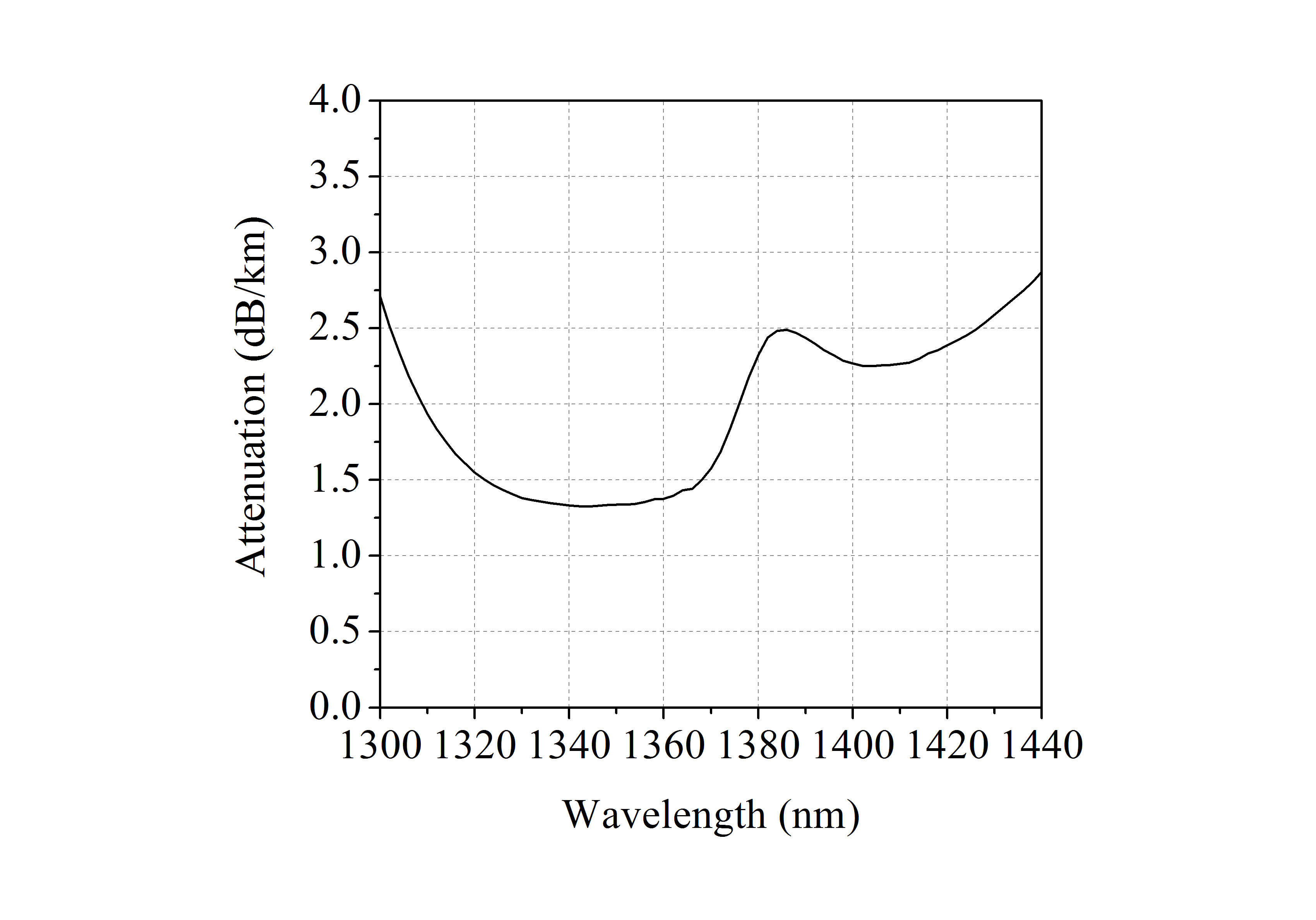}
\caption{Attenuation spectrum measured on a 233~m sample of Ho-doped fiber (fiber \#1) close to 1380~nm, in order to estimate the concentration of OH groups.}
\label{Fig_OH_loss_iXblue}
\end{figure}

According to the values reported in \cite{Humbach_1996}, the linear attenuation relative to OH concentration is about 60~dB/km/ppm close to 1380~nm, wavelength of the first overtone of an OH fundamental stretching vibration. We measure the linear attenuation spectrum of a 233~m sample of fiber \#1 around this wavelength, using the usual cut-back method. The result of this measurement is presented on Figure \ref{Fig_OH_loss_iXblue}. As expected, a distinct peak appears at 1380~nm, showing an attenuation of about 2.5~dB/km. The corresponding OH concentration can be estimated to about 0.04~ppm, and the contribution of OH to the attenuation at the laser wavelength is lower than 1~dB/km. This value is remarkably low compared to the 1~ppm reported in \cite{Simakov_2013} or even the previous record low value of 0.14~ppm reported in \cite{Friebele_2014}. As discussed in section \ref{Sec_discussion}, such a glass purity is an important requirement in the race for efficient fiber lasers, in this spectral domain.

\subsection{Pair induced non-saturable absorption}

Another notable contribution to the attenuation of pump and signal is due to the formation of rare-earth ion pairs in highly doped fibers \cite{Kurkov_2010}. The electronic structure of the holmium ion, like in erbium or thulium, allows transfers of energy between ions in close vicinity. The pair induced quenching (PIQ) typically alters the pumping efficiency of a fiber laser, and can also affect the transmission of the amplified signal \cite{Delevaque_1993}. When the pump or signal power is very high compared to the saturation power of the corresponding transition, one of the ions of a pair in the excited state decays non-radiatively to the ground state while the other is up-converted to a higher energy level. This non-saturable absorption thus translates to a loss mechanism for both signal and pump beams.

\begin{figure}[h!]
\centering\includegraphics[width=13.5cm, trim={.3cm 0cm 0 0}, clip]{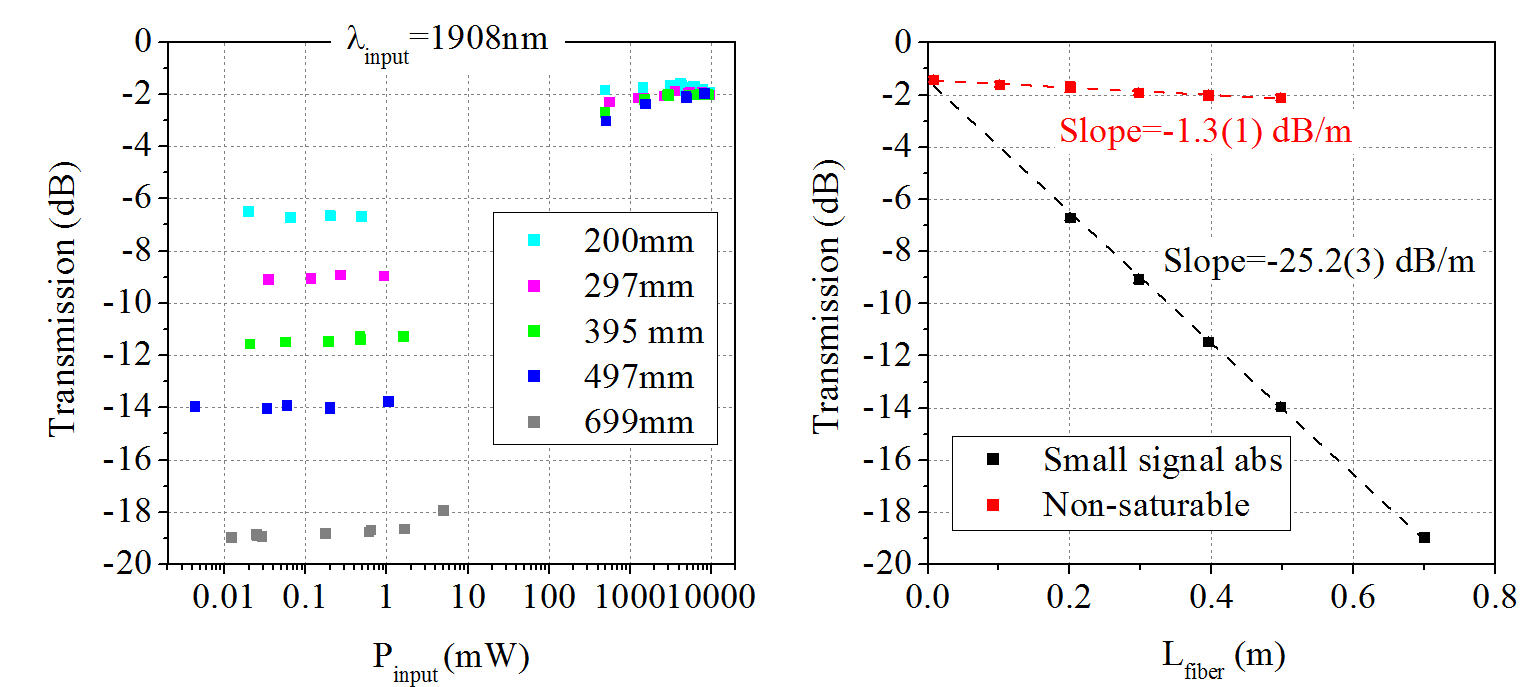}
\caption{Left: measurement of the transmission as a function of the input power $P_\text{in}$ at $\lambda_\text{in}=1908$~nm for various length of the FUT. Right: Evolution of the small-signal ($P_\text{in}<1$~mW) and non-saturable ($P_\text{in}>1$~W) transmissions as a function of the FUT length.}
\label{Fig_nonsat_abs}
\end{figure}

The non-saturable absorption is measured by varying the power injected in the FUT core at $\lambda_\text{in}=1908$~nm, following a standard method \cite{Boivin_2010}. The probe source is a Tm-doped fiber laser (IPG TLR10-1908-LP). The nearly gaussian beam is coupled into a short piece of single-mode PM1950 fiber. About 70\% of the power of the fiber laser probe can be injected into the PM1950 fiber, which is fused to a sample of fiber \#1. The initial length of the sample is about 0.7~m. For input powers lower than 10~mW, the output power is measured with a pyroelectric detector, and we use a calorimeter for input powers higher than a few hundreds of~mW.

We report on Fig.~\ref{Fig_nonsat_abs} (left) the transmission as a function of the input power, for several length $L_\text{FUT}$ of fiber \#1. This figure clearly illustrates the two regimes of small-signal transmission $T_\text{Ho}$ (for $P_\text{input}<1$~mW) and saturated transmission $T_\text{sat}$ (for $P_\text{input}>1$~W). Obviously these raw transmission values are biased with various losses, which are independent of the FUT length (transmissions of fiber components, splices at the input/output of the FUT...). By plotting the asymptotic values $T_\text{Ho}(L_\text{FUT})$ and $T_\text{sat}(L_\text{FUT})$ as a function of $L_\text{FUT}$ (Fig.~\ref{Fig_nonsat_abs} right), we deduce from the slopes of the linear fits the corresponding small-signal absorption $\alpha_\text{Ho}=-T_\text{Ho}$ and non-saturable absorption $\alpha_\text{ns}=-T_\text{ns}$ of holmium in this sample. The constant loss contribution corresponds to the intercept at $L_\text{FUT}=0$. In the case of fiber \#1, we find $\alpha_\text{Ho}(\lambda_\text{in})=25.2 \pm 0.3$~dB/m and $\alpha_\text{ns}(\lambda_\text{in})=1.3 \pm 0.1$~dB/m.

These two values are related to the fraction $2k$ of ions in pairs by the expression derived in \cite{Myslinski_1997}:
\begin{equation}
\label{eq_abs_nonsat}
\alpha_\text{ns}(\lambda)=2k\cdot\alpha_\text{Ho}(\lambda) \cdot \left[1- \dfrac{\sigma_\text{abs}(\lambda)+\sigma_\text{em}(\lambda)}{2\sigma_\text{abs}(\lambda)+\sigma_\text{em}(\lambda)} \right]
\end{equation}
where $\sigma_\text{abs}(\lambda)$ and $\sigma_\text{em}(\lambda)$ are the absorption and emission cross-section spectra reported above (see Fig.~\ref{Fig_Ho_spectra_setup}). Considering the cross-sections at $\lambda_\text{in}=1908$~nm ($\sigma_\text{abs}=1.9 \ 10^{-25}\text{m}^2$ and $\sigma_\text{em}=1.6 \ 10^{-25}\text{m}^2$), we find that $2k= 15\pm 1\%$. Here again, the literature is very scarce about direct measurements of ion pairs concentration. Our value is comparable with the very few measurements reported on holmium doped fibers with similar holmium concentration \cite{Kurkov_2010, Wang_2019}.

\section{Realization and analysis of a high power laser}

Since the general objective of this work is the ultimate efficiency and maximum power of a all-fiber holmium laser, we report in the following section the study of the laser oscillator configuration. Future works will be dedicated to the single-frequency amplifier configuration. Both experimental results and simulation analysis of the fiber laser are presented.

\subsection{Fiber laser setup}
\label{Sec_setup}

\begin{figure}[h!]
\centering\includegraphics[width=13cm, trim={0cm 0cm 0 0}, clip]{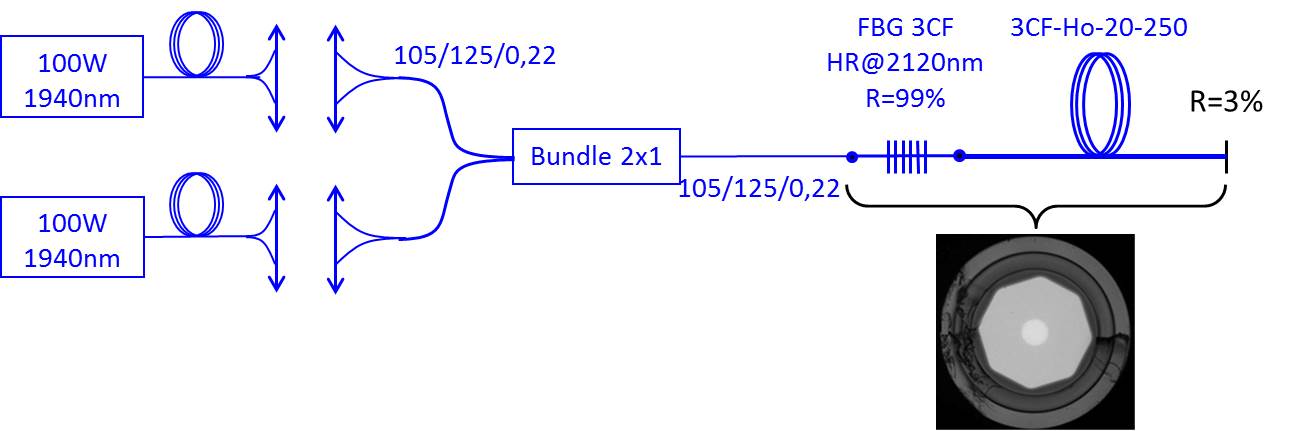}
\caption{Design of the holmium-doped fiber laser configuration. The picture inserted illustrates the triple-clad structure of the active fiber (3CF-Ho). The passive fiber of the FBG is also a 3CF, with a 250~\textmu m diameter circular cladding.}
\label{Fig_setup_ONERA}
\end{figure}

Our setup illustrated on Fig.~\ref{Fig_setup_ONERA} is a standard oscillator configuration, based on a LMA triple-clad holmium doped fiber (IXF-3CF-Ho-O-20-250-300-0.08, see insert in Fig.~\ref{Fig_setup_ONERA}). The triple clad (3CF) structure, also named all-glass cladding, is now familiar in works on in-band pumped fibers emitting in the 2~\textmu m spectral range \cite{Hemming_2012,Friebele_2014,Shardlow_2016}. The active core (20~\textmu m diameter) is surrounded by a Ge-doped pedestal (64~\textmu m diameter) that guarantees a low numerical aperture of 0.08. The pump is guided in the pure silica cladding (250~\textmu m diameter flat to flat), thanks to a surrounding layer of fluorine (F) doped silica (300~\textmu m diameter). This interface is shaped as an octagon, in order to break the angular symmetry and force the pump beam to cross the doped core. Due to the limitation in F concentration, the numerical aperture of the pump guiding interface is limited to 0.22. The mechanical protection of the fiber is ensured by a polymer cladding, which results in a total external diameter of 450~\textmu m.

The absorption by the holmium ions in the 3CF is measured by a cut-back method on a sample of initial length 1.5~m. We succeed in keeping the fraction of light coupled into the pedestal to less than 1\%, and find a core absorption of 86$\pm$3~dB/m at the peak absorption wavelength 1946~nm for the fundamental guided mode. The ion concentration was estimated otherwise to about $5.3\ 10^{25}\ \text{m}^3$, from the quantity of holmium oxyde introduced for the preform fabrication. Using eq. (\ref{eq_absorption}) and taking into account the overlap $\Gamma=84$\% between doped core and optical mode, this estimation yields an absorption $\alpha_\text{Ho}^\text{peak} \simeq 20\ \text{m}^{-1}=87~$dB/m, which is consistent with the value measured directly in the fiber for the fundamental mode.



The highly reflective mirror is a fiber Bragg grating (FBG) written in a passive 3CF fiber with the same dimensions and NA as the active fiber, except that this clad is circular instead of octagonal. The FBG is manufactured by iXblue using a proprietary process that allows a minimal heating at high signal powers close to 2~\textmu m. Its reflectivity is about 99\% at $\lambda_s$, with a 3~dB linewidth of 2~nm. The laser cavity output coupler is obtained by the flat cleaved end facet of the active fiber. The Fresnel reflectivity of this silica-air interface is approximately 3\%. 

The pump light is provided by two commercial fiber lasers delivering collimated single-mode beams (M$^2$<1.05) with a power up to 120~W each at 1940~nm, in a linewidth lower than 1~nm. The two beams are focused through AR coated dry-silica lenses (Corning 7979) into the multimode fibers (MMF: 105/125/0.22) of a 2x1 pump combiner. The output port of the combiner is an identical 105/125/0.22 fiber. It is then spliced to the passive 3CF with the core-engraved FBG (pump diameter 250~\textmu m, silica diameter 300~\textmu m), which is itself spliced to the 3CF-Ho active fiber (pump diameter 250~\textmu m, silica diameter 300~\textmu m). Between the free-space output of the commercial pump lasers, and the output of a short piece (about 5cm) of the 3CF-Ho fiber, we measured a transmission of 93$\pm$1\%. This value corresponds to the following sequence for the pump beams: focusing lens / injection in the input MMF of the combiner / 2x1 combiner / {combiner MMF output - FBG} splice/ {FBG - Ho fiber} splice.

The 3CF-Ho fiber is cooled by a small fan blowing above. The fiber output beam, carrying the laser signal and the residual pump, is collimated with another dry-silica lens. The material and AR-coating of the lenses allows a transmission higher than 99\% at the pump and signal wavelengths. A dichroic filter transmits the residual pump (T=95\% at $\lambda_p$) and reflects the laser signal (R=99\% at $\lambda_s$). The output power at both wavelength is measured by calorimeters on each side of the filter. A fraction of the signal port is also coupled into an optical spectrum analyzer. 

\subsection{Experimental results}
\label{Sec_exp_results}

\begin{figure}[h!]
\centering\includegraphics[width=14cm, trim={.5cm .3cm 0cm .3cm}, clip]{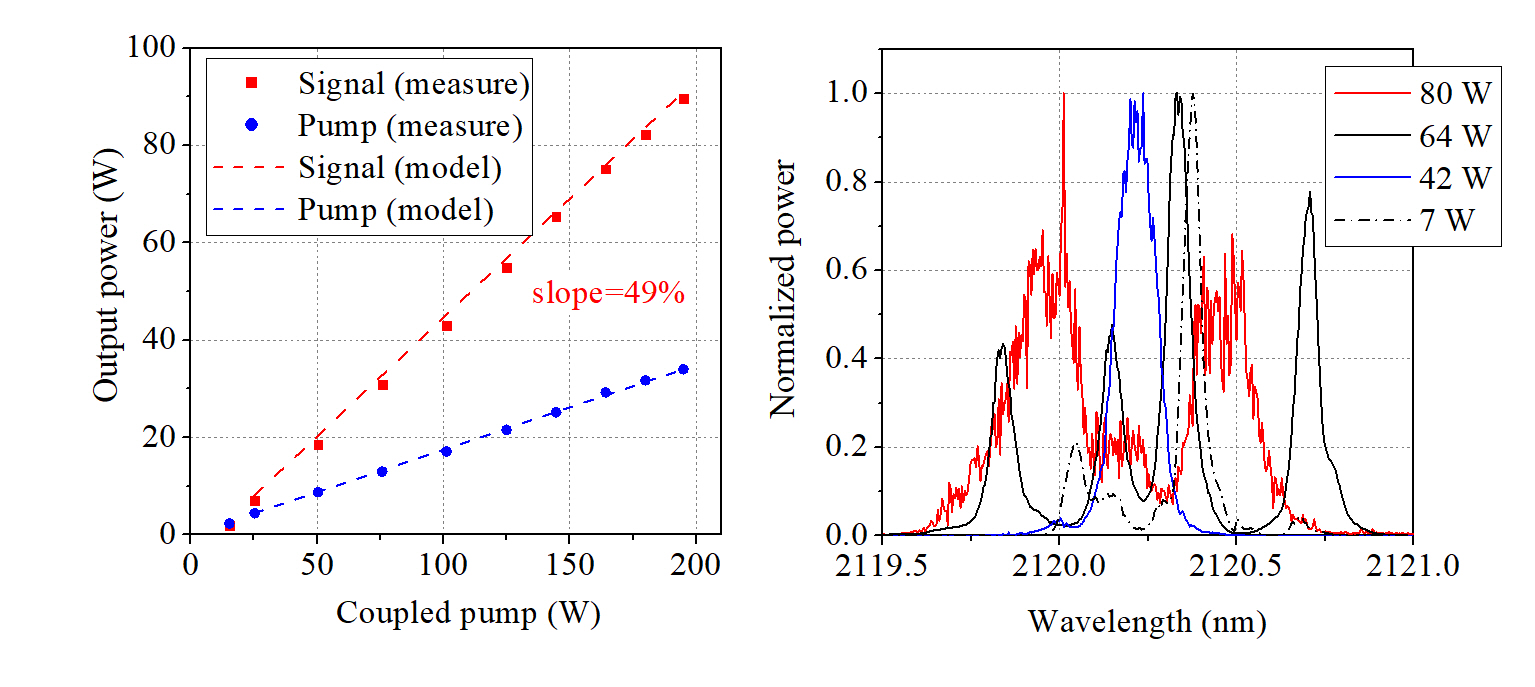}
\caption{Left: evolution of the output signal and residual pump powers (respectively red and blue), as a function of the pump power. The dots represent the experimental data, the lines the result of our simulation. Right: Signal spectra for various values of P$_\text{sig}$, for a resolution 0.05~nm.}
\label{Fig_laser3CF}
\end{figure}

Using our numerical model described below, with the inputs obtained from the measurements listed in Section \ref{Sec_spectro}, we first estimated the fiber length required to maximize the signal power to about L$\simeq$16~m. 
We first tested the laser efficiency with this fiber length. The evolution of the signal power as a function of the pump power is shown on Fig.~\ref{Fig_laser3CF} (left).

For the maximum coupled pump power of about 200~W, the laser delivers an output power of 90~W at 2120~nm. The optical efficiency, defined as the slope of the signal power as a function of the coupled pump power, reaches 49\%. As a comparison, the maximum output power reported at this wavelength from a Ho fiber laser in the same configuration is about 400~W, for a 1~kW pump power \cite{Hemming_2013}. Up to a pump power of 400~W, the laser efficiency was slightly lower than 50\%. 

The efficiency that we report is identical, despite a fundamental difference. Indeed, the overlap between the pump beam and the active region in our fiber is much less favourable. The Ho-doped fiber presented here results from the first trials of triple clad manufacturing at iXblue. Due to the difficulties of filling properly a big preform core and machining a small preform, the clad diameter of our fiber could not be reduced to less than 250~\textmu m, instead of 112~\textmu m for the Nufern fiber of the DSTG \cite{Simakov_2013}. Therefore the pump/core overlap in our fiber is lower by more than a factor of 4, which reduces the thermal load. The corollary drawback is that for the same Ho concentration, the fiber length must be much longer, thus increasing the loss on the pump and signal fields and reducing the laser efficiency. The lower OH concentration and pair fraction of our fiber allowed however a state of the art efficiency. As described in the next section, this performance is well understood in light of our numerical model.

The laser spectrum was monitored as the laser signal power was increasing. As illustrated on Fig.~\ref{Fig_laser3CF} (right), the bandwidth remained in the range of 1~nm, as defined by the FBG, up to the maximum laser power. More importantly, one can see that the center wavelength does not increase, as could be the case for example if the thermal load expands the grating. As for the shape of the spectrum at 80~W, the 0.05~nm resolution of the OSA allows the observation of a structured spectrum. Numerous longitudinal modes oscillate in the long fiber cavity, and the result of their competition can vary rapidly. The structured shape may result from variations of the spectrum on a time scale close to that of the OSA. The study of the spectrum shape is out of the scope of this paper but will be investigated in later works.

Finally, the laser was tested for a fiber length of 10~m, in order to check the influence on the efficiency. In that case, the laser power was 70~W at 2120~nm, for a pump power of 170~W, and the optical efficiency was 43\%. These values are also in good agreement with the model presented below.

\subsection{Comparison with numerical modeling}
\label{Sec_simu_results}

The powers at the laser signal and pump wavelength are calculated along the fiber path by solving the usual rate and propagation equations \cite{Desurvire_EDFA, Wang_2018}, with the common approximation of the effective overlap integral between the active region and the propagating mode to suppress the radial dependence \cite{Giles_1991}. We describe the holmium amplifying medium as a simple homogeneously broadened two-level system, considering only the $^5I_8-{^5I_7}$ transition.

The numerical model takes into account the physical parameters of the fiber (core/clad diameters and numerical apertures), as well as the data that we measured and reported above (interaction cross-sections, fluorescence lifetime, background loss, absorption, losses...). The main values required for the simulation are listed in Table \ref{tab_Fiber_parameters}.

\begin{table}
\centering
\caption{List of the parameters of the Ho-doped fiber laser used in the numerical simulation}
  \label{tab_Fiber_parameters} 
\begin{tabular}{|c|c|c|}
  \hline
  \rowcolor{lightgray}
  Parameter & Value & Comment \\
  \hline
  $\varnothing_\text{core}$ & 20~\textmu m & Core diameter\\
  $\varnothing_\text{clad}$ & 250~\textmu m & Clad diameter\\
  $\text{NA}_\text{core}$ & 0.08 & Numerical aperture\\
  $\tau_0$ & 0.7 ms & Laser transition fluorescence lifetime  \\
  $\alpha_\text{Ho}^\text{peak}$ & 86~dB/m & Ho$^{3+}$ peak absorption at 1946~nm \\
  $2k$ & 23\% & Fraction of Ho$^{3+}$ ions in pairs \\
  $\zeta_\text{clad}(\lambda_p)$ & 4 dB/km & Pump attenuation in the clad  (mainly multi-phonon) \\
  $\zeta_\text{core}(\lambda_s)$ & 40 dB/km & Signal attenuation in the core (mainly multi-phonon) \\

  \hline
\end{tabular}
\end{table}

The fraction of ions in pairs $2k$ could not be measured directly on the 3CF-Ho LMA fiber, but we can estimate it from the value that was measured in the fiber \#1 ($2k=15\%$ for  $\alpha_\text{Ho}^\text{peak}=57$~dB/m). Assuming a proportional variation around these absorption values, we expect $2k=23\%$. The effect of PIQ on the laser efficiency is modeled in a simple way, since the powers involved are much higher than the saturation power $P_\text{sat}=h\nu A_\text{core}/(\sigma_\text{abs}+\sigma_\text{em})\Gamma\tau_0$, at both $\lambda_p$ and $\lambda_s$ (about 40~mW and 260~mW respectively). The attenuation induced by the ion pairs is thus essentially constant along the fiber and can be identified to the non-saturable absorption (eq.~\ref{eq_abs_nonsat}). For the signal propagating in the core, the contribution $\alpha_\text{ns}(\lambda_s)$ is added directly to the attenuation $\zeta_\text{core}(\lambda_s)$, while for the pump propagating in the clad, the term $\alpha_\text{ns}(\lambda_p)$ must be weighted by the surface ratio $A_\text{core}/A_\text{clad}$ between core and clad before addition to $\zeta_\text{clad}(\lambda_p)$.

The only parameters that could not be derived from our measurements on the Ho-doped fiber are the contributions to the silica attenuation from the multi-phonon absorption $\zeta_\text{silica}$ by the SiO$_4$ bonds, at the signal and pump wavelengths. Therefore we use these values as parameters to fit the numerical simulation to the experimental result: $\zeta_\text{silica}(\lambda_s)$ is used to fit the output signal power, and $\zeta_\text{silica}(\lambda_p)$ the residual output pump power. The results of the simulation for the signal and pump powers at the fiber output are displayed as the dotted lines (red and blue respectively) on Fig.~\ref{Fig_laser3CF} (left), as a function of the coupled pump power. The best fit between measurement and simulation for the output signal is obtained for $\zeta_\text{silica}(\lambda_s)=40$~dB/km and $\zeta_\text{silica}(\lambda_p)=4$~dB/km. These values are of the same order as those reported in the literature \cite{Izawa_1977,Garrett_1982}.

We note that the value of $2k$ estimated for our fiber, and validated by our model, is notably lower than the value $2k=30\%$ inferred in \cite{Wang_2018} for the fiber tested in \cite{Hemming_2013}, with a lower peak absorption of 70~dB/m. However the direct comparison is not relevant, since the fraction of paired ions is know to be highly dependent on the host composition \cite{Arai_1986,Kiritchenko_2015,Savelii_2017, Wang_2019}. The evolution of $2k$ as a function of the Ho$^{3+}$ concentration in our silica hosts is the subject of an ongoing work.

\section{Discussion on the power limitation}
\label{Sec_discussion}
\begin{figure}[h!]
\centering\includegraphics[width=9cm, trim={1cm .5cm 1cm .5cm}, clip]{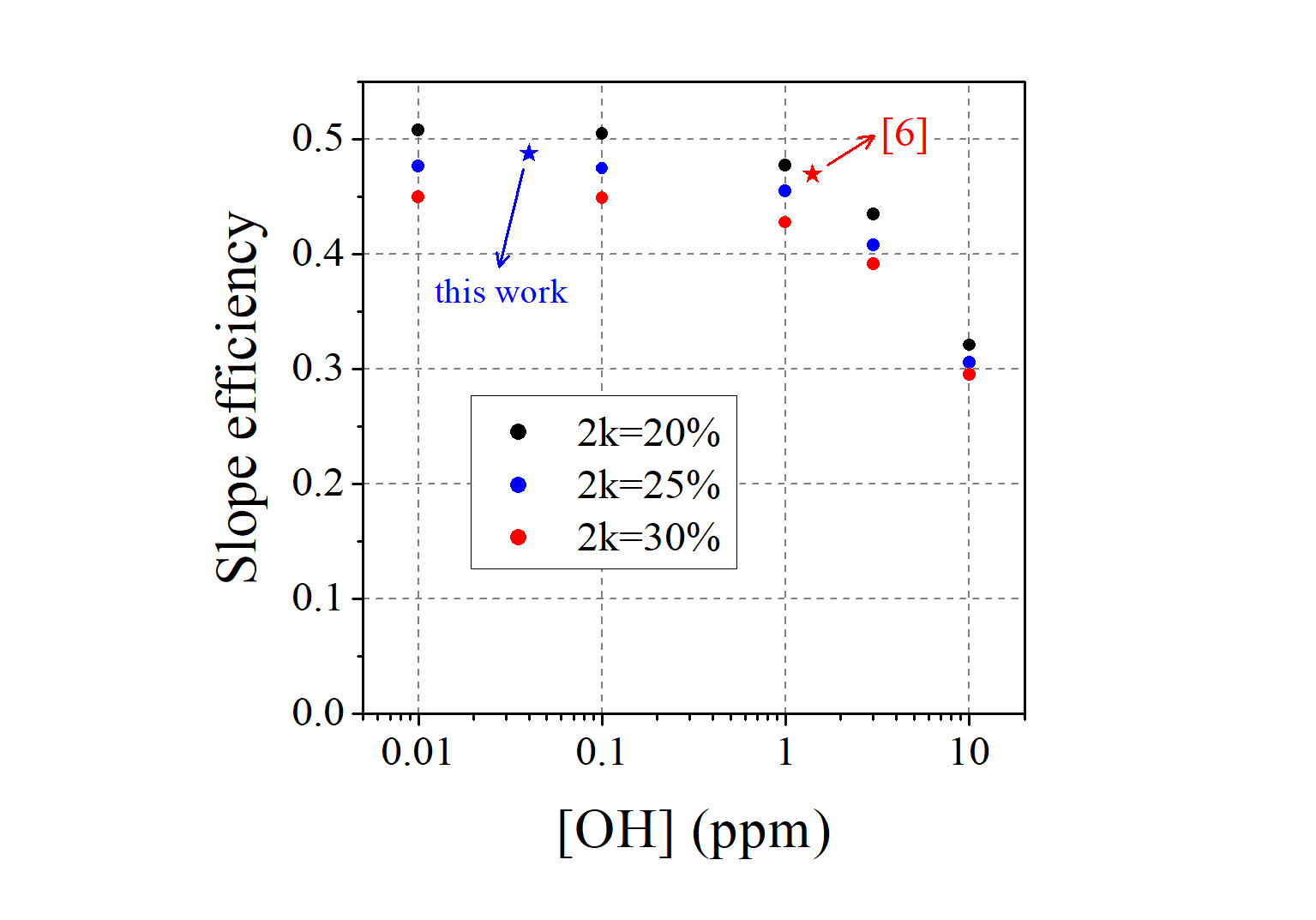}
\caption{Influence of the concentration of OH groups on the efficiency of a fiber laser, with the parameters listed in Table \ref{tab_Fiber_parameters}, for various values of fraction $2k$ of Ho$^{3+}$ ions in pairs. The blue and red stars represent respectively the efficiencies of our fiber laser, and of the laser reported in \cite{Hemming_2013} (for $P_\text{signal}$ up to $\simeq 200~W$).}
\label{Fig_influence_2k_OH}
\end{figure}

We now use our model to illustrate the role of the two main causes of attenuation in the Ho-doped core, namely OH groups and holmium pairs. For holmium doped fibers, the literature shows that the concentration [OH] ranges typically between 0.01~ppm and 10~ppm. As for the fraction of ions in pair $2k$, it can reach 30\% \cite{Wang_2018}, and we suppose that it could be reduced down to 20\%. Considering a fiber with the dimensions and parameters listed in Table \ref{tab_Fiber_parameters}, the simulation is run for various values of [OH] and $2k$. For each pair of these parameters, we calculate the maximum laser slope efficiency, looking for the optimal fiber length (Fig.~\ref{Fig_influence_2k_OH}).

We find that for a given value of $2k$, reducing [OH] from 1~ppm to 0.1~ppm improves the slope efficiency by less than 3\% percent, while the same improvement can be obtained by reducing $2k$ from 30\% to 25\%. This first observation shows that for [OH] lower than 1~ppm, the main effort should lie rather on avoiding the ion pairs than on reaching a dryer silica core. The dilution of the Ho$^{3+}$ ions hence appears as the main concern in the design of future fibers.

Finally, we used our simulation tool to study the holmium doped fiber that still holds the highest laser power published to date \cite{Hemming_2013}, using the values gathered in \cite{Wang_2018}. For the contribution of the ion pairs, we consider again that it is reduced to the non-saturable absorption, using the value $2k=30\%$ that was inferred from the comparison between model and experimental results \cite{Wang_2018}. With this approximation, we find for example $P_s=183$~W for $P_p=400$~W, and a slope efficiency of 47\%, both values in excellent agreement with the experimental data. The agreement suggests that our model, with the approximations mentioned above, yields the same accuracy as the complete model presented in \cite{Wang_2018,Wang_2019}.


\section{Conclusion}
We report here on the study of a holmium doped fiber laser for the high power continuous regime, using triple-clad fibers adapted to the in-band pumping in the 2~\textmu m band. One interest of our work lies in the detailed spectroscopy of the laser transition of the holmium ion in a silica host: lifetime, interaction cross-section spectra, fraction of Ho$^{3+}$ pairs... We also demonstrate a laser efficiency of about 50\%, which corresponds to the present state of the art for a continuous laser in the 100~W class. Another major aspect of this study is the comparison with the numerical simulation that uses the values obtained from the spectroscopic measurements. The agreement of our simulation with the experiment shows that it provides a good understanding of the laser system. It will thus be useful to design more efficient Ho-doped fibers for high power laser sources emitting at $\lambda>2.1$~\textmu m. Further work is now in preparation to improve the laser efficiency and describe more precisely the role of PIQ.


\section*{Acknowledgments}
The authors are thankful to the french defense direction DGA for financial support.

\section*{Disclosures}
The authors declare no conflicts of interest.


%




\bibliography{2019_LaserHo_OpEx_revised}

\end{document}